\documentclass[onecolumn,preprintnumbers,eqsecnum,superscriptaddress,12pt]{revtex4}
\usepackage{amsfonts}
\usepackage{amssymb}
\usepackage{amsmath}
\usepackage{epsf}
\usepackage{graphicx}
\usepackage{dcolumn}
\usepackage{bm}

\usepackage{color}

\def\nn{{\nonumber}}

\newcommand{\beq}{\begin{equation}}
\newcommand{\eeq}{\end{equation}}
\newcommand{\beqs}{\begin{eqnarray}}
\newcommand{\eeqs}{\end{eqnarray}}

\newcommand{\be}{\begin{equation}}
\newcommand{\ee}{\end{equation}}
\newcommand{\bea}{\begin{eqnarray}}
\newcommand{\eea}{\end{eqnarray}}

\begin{document}

\title{The first order hydrodynamics via AdS/CFT correspondence in the Gauss-Bonnet gravity }
\author{Ya-Peng Hu}\email{yapenghu@itp.ac.cn}
\address{
Center for High-Energy Physics, Peking University, Beijing 100871, China}
\address{Key Laboratory of Frontiers in Theoretical Physics,
Institute of Theoretical Physics, Chinese Academy of Sciences, P.O. Box 2735, Beijing 100190, China}

\author{Huai-Fan Li}\email{huaifan99@yahoo.com.cn}
\address{Key Laboratory of Frontiers in Theoretical Physics,
Institute of Theoretical Physics, Chinese Academy of Sciences, P.O. Box 2735, Beijing 100190, China}
\address{Department of Physics and Institute of Theoretical Physics, Shanxi Datong University, Datong 037009, China}
\address{Department of Applied Physics, Xi'an Jiaotong University, Xi'an 710049, China}

\author{Zhang-Yu Nie}\email{niezy@itp.ac.cn}
\address{Key Laboratory of Frontiers in Theoretical Physics,
Institute of Theoretical Physics, Chinese Academy of Sciences, P.O. Box 2735, Beijing 100190, China}

\begin{abstract}
In the spirit of the AdS/CFT correspondence, we investigate the
hydrodynamics of the dual conformal field in the Gauss-Bonnet
gravity. By considering the parameters of the boosted black brane in the Gauss-Bonnet gravity as functions of boundary coordinates, and then solving the corresponding correction terms, we calculate the first order stress-energy tensor of the dual
conformal field. From this first order stress-energy tensor, we also
obtain the shear viscosity and entropy density. And these results
are consistent with those of some previous works from the effective
coupling of gravitons.
\end{abstract}

\maketitle

\vspace*{1.cm}

\newpage

\section{Introduction}
The AdS/CFT
correspondence~\cite{Maldacena:1997re,Gubser:1998bc,Witten:1998qj,Aharony:1999ti}
provides a theoretical way to understand the strongly coupled field
theory. It has been applied to study various of physical problems such as
the holographic superconductor~\cite{Herzog:2009xv}, and has became a quite useful tool to extract the hydrodynamic properties of the strongly coupled conformal field in the quantum chromodynamics (QCD).

It is well-known that any interacting quantum field theory can be
effectively described by hydrodynamics in the long wavelength limit.
Using AdS/CFT, the shear viscosity of the boundary fluid has been
calculated in theories dual to Einstein
gravity~\cite{Policastro:2001yc,Kovtun:2003wp,Buchel:2003tz,Kovtun:2004de}.
And the ratio of shear viscosity over entropy density $\eta/s$ is
found to be $1/4\pi$ for a large class of conformal field theories
with gravity dual in the large N
limit~\cite{{Mas:2006dy},{Son:2006em},{Saremi:2006ep},{Maeda:2006by},{Cai:2008in}}.
Then it is found that the ratio $\eta/s$ only depends on the value
of effective coupling of transverse gravitons evaluated on the
horizon~\cite{Brustein:2008cg,Iqbal:2008by,Cai:2008ph,Astefanesei:2010dk}. Note that,
this ratio gets only positive corrections from some Large N
effect~\cite{Buchel:2004di,Myers:2008yi}. Therefore, this value is
once conjectured as an universal lower bound for all
materials~\cite{
Policastro:2002se,Buchel:2004qq,Cohen:2007qr,Son:2007vk,Cherman:2007fj,Chen:2007jq,Fouxon:2008pz,Dobado:2008ri,Landsteiner:2007bd}.
However, the later study showed that the higher derivative gravity
such as Gauss-Bonnet(GB) gravity can modify the lower bound to a
lower
value~\cite{Brigante:2007nu,Brigante:2008gz,KP,Brustein:2008cg,Buchel:2009sk,deBoer:2009pn,Camanho:2009vw}.
And the case dual to general LoveLock gravity can give further
modification to this lower bound~\cite{deBoer:2009gx,Camanho:2009hu}.

Recently, the study of the hydrodynamics via dual gravity has been
further developed as the Fluid/Gravity
correspondence~\cite{Bhattacharyya:2008jc}. This correspondence can
provide a more systematic way that map the boundary fluid to the
bulk gravity. As we know, the hydrodynamical equations can be
governed by the conservation of energy and other conserved global
charges. In the relativistic case, these can be expressed by the
conservation of the stress-energy tensor $T^{\mu\nu}$ and the charge
currents $J^{\mu}$, and finally can be completely characterized by
the thermodynamic variables (such as the energy density $\rho$ and
pressure $P$) and the velocity field of fluid $u^{\mu}$. When the
fluid is not ideal, some parameters such as the shear viscosity
$\eta$ would appear to characterize the effects of dissipation of
fluid. On the other hand, viewed from the side of the dual gravity,
the effects of dissipation of the fluid should also be constructed
from the bulk gravity. In Reference~\cite{Bhattacharyya:2008jc}, the
authors indeed constructed the stress-energy tensor of the non-ideal
fluid order by order from the bulk solution in Einstein gravity.
From the first order stress-energy tensor, the shear viscosity was
calculated, and the ratio of the shear viscosity over entropy
$\eta/s$ was same as the previous result $1/4\pi$. Following this
method, some more complicated cases such as the existence of
electromagnetic field at the boundary are also
considered~\cite{Hur:2008tq,Erdmenger:2008rm,Banerjee:2008th,Tan:2009yg}.
Their results are consistent with the previous results, and the
ratios are still $1/4\pi$. Therefore, it will be interesting to
compute the non-trivial ratio $\eta/s$ in Gauss-Bonnet gravity by
using this new method and make a comparison to the previous results.
Note that, following the reference~\cite{Bhattacharyya:2008jc}, the
higher stress-energy tensor of the dual conformal theory on the
boundary can be calculated order by order from the bulk solution.
But we just take its shear viscosity and entropy density into
account, therefore, the first order stress-energy tensor are enough
to extract these information. Our results show that the shear
viscosity and entropy density are consistent with the previous
results, which can also be considered as a good test of this
Fluid/Gravity correspondence with a more general gravity in the
bulk.

The following of the paper is scheduled as follows. In Sec.~II and Sec.~III, we shortly review some properties of the Gauss-Bonnet gravity and the boosted black brane solution, respectively. In Sec.~IV, we construct the first order stress-energy tensor of the conformal field via the dual gravity in the $5$-dimensional spacetime case. Sec.~V is devoted to the conclusion and discussion.


\section{The Gauss-Bonnet gravity and field equations}
The action of the $d$ dimensional Gauss-Bonnet gravity with a negative cosmological constant $\Lambda=-(d-2)(d-1)/2\ell^2$ can be
\begin{eqnarray}
\label{action} I=\frac{1}{16 \pi G}\int_\mathcal{M}~d^dx \sqrt{-g}
\left(R-2 \Lambda+\alpha L_{GB} \right),
\end{eqnarray}
where $R$ is the Ricci scalar, $\alpha$ with dimension $(length)^2$ is
the GB coefficient and the GB term $L_{GB}$ is
\begin{eqnarray}
\label{LGB} L_{GB} =R^2-4R_{\mu \nu}R^{\mu \nu}+R_{\mu \nu \sigma
\tau}R^{\mu \nu \sigma \tau}.
\end{eqnarray}
Note that, the GB term is a topological invariant in four dimensional spacetimes. And the Gauss-Bonnet gravity can be viewed as the second order term in the Lovelock gravity which preserves the property that the equations of motion involve
only second order derivatives of the metric as that in the Einstein gravity~\cite{Love}.

The generalized surface term of the GB gravity is~\cite{Myers:1987yn,Brihaye:2008xu}
\begin{equation}
I_{\mathrm{sur}}=-\frac{1}{8\pi G}\int_{\partial
\mathcal{M}}d^{d-1}x\sqrt{-\gamma }K-\frac{\alpha}{4\pi G}\int_{\partial \mathcal{M}}d^{d-1}x\sqrt{-\gamma }%
 \left( J-2{\rm G}_{ab} K^{ab}\right) ~, \label{Surfaceterm}
\end{equation}
where $\gamma _{ab }$ is the induced metric on the boundary, $K$ is
the trace of the extrinsic curvature of the boundary, ${\rm G}_{ab}$ is the Einstein tensor of the metric $\gamma _{ab}$ and $J$ is the
trace of the tensor
\begin{equation}
J_{ab}=\frac{1}{3}%
(2KK_{ac}K_{b}^{c}+K_{cd}K^{cd}K_{ab}-2K_{ac}K^{cd}K_{db}-K^{2}K_{ab})~.
\label{Jab}
\end{equation}
From~(\ref{Surfaceterm}), it is obvious that the first term is just the well-Known Gibbons-Hawking surface term and the second term is the counterpart for the GB gravity.

Through the variation of the action (\ref{action}) with respect to the
bulk metric, we can obtain the equation of the Gauss-Bonnet gravity
\begin{eqnarray}
\label{eqs} R_{\mu \nu } -\frac{1}{2}Rg_{\mu \nu}+\Lambda g_{\mu \nu
}+\alpha H_{\mu \nu}=0~,
\end{eqnarray}
where
\begin{equation}
\label{Hmn}
H_{\mu \nu}=2(R_{\mu \sigma \kappa \tau }R_{\nu }^{\phantom{\nu}%
\sigma \kappa \tau }-2R_{\mu \rho \nu \sigma }R^{\rho \sigma
}-2R_{\mu
\sigma }R_{\phantom{\sigma}\nu }^{\sigma }+RR_{\mu \nu })-\frac{1}{2}%
L_{GB}g_{\mu \nu }  ~.
\end{equation}

\section{Boosted black brane in the Gauss-Bonnet gravity}
Although it is difficult to find exact solutions of the GB equations due to its nonlinearity and complexity, some exact solutions have already been founded~\cite{Des,Whee,Myers,Cai1,Cai:2001dz}. Here we are interested in is the $5$-dimensional black brane solution~\cite{Cai:2001dz}
\begin{eqnarray}
ds^2=\frac{dr^2}{r^2f(r)}+\frac{r^2}{\ell_c^2}
 \left(\mathop\sum_{i=1}^{3}dx_i^2 \right)-r^2f(r) dt^2, \label{Solution}
\end{eqnarray}%
where
\begin{eqnarray}
\label{f-BH}
 f(r) = \frac{1}{4\alpha}
 \bigg (
 1-\sqrt{1-\frac{8\alpha}{\ell^2}(1-\frac{b^4}{r^{4}})}~
 \bigg).
\end{eqnarray}%
Note that, as  $r \to \infty$, $f(r)$ approximates ${1/\ell_c^2}$.
Here $\ell_c$ is the effective radius of the AdS spacetime in GB gravity, and its expression in $5$-dimensional case is
\begin{eqnarray}
\label{lc}
\ell_c=\ell\sqrt{\frac{1+U}{2}},~~{\rm~~with~~~~}U=\sqrt{1-\frac{8\alpha}{\ell^2}}.
\end{eqnarray}
From which, we can also find
the existence of an upper bound for the GB coefficient, $\alpha\leq
\alpha_{max}=\ell^2/8$, which holds for all $5$-dimensional asymptotically
AdS solutions in GB gravity.

From~(\ref{Solution}), we easily find that the horizon of the black brane is located at $r=b$. The Hawking temperature and entropy density are
\begin{eqnarray}
T&=&\frac{(r^2f(r))'}{4 \pi}|_{r=b}=\frac{b}{\pi},\label{Temperature}\\
s&=&\frac{b^3}{4G \ell_c^3}.\label{entr}
\end{eqnarray}

In the Eddington-Finkelstin coordinate system, the solution~(\ref{Solution}) can be rewritten as
\begin{equation}
ds^2 = - r^2 f(r)dv^2 + 2 dv dr + \frac{r^2}{\ell_c^2}(dx^2 +dy^2 +dz^2). \label{Solution1}
\end{equation}
where $v=t+r_*$ with $dr_*=dr/(r^2f)$.
The boosted black brane can be obtained from~(\ref{Solution1}) via a coordinate transformation which is generated by a subalgebra of the isometry group of $AdS_{5}$. And its line element is
\begin{equation}\label{rnboost}
ds^2 = - r^2 f(r)( u_\mu dx^\mu )^2 - 2 u_\mu dx^\mu dr + \frac{r^2}{\ell_c^2}P_{\mu \nu} dx^\mu dx^\nu,
\end{equation}
with
\begin{equation}
u^v = \frac{1}{ \sqrt{1 - \beta_i^2} }~~,~~u^i = \frac{\beta_i}{ \sqrt{1 - \beta_i^2} },~~P_{\mu \nu}= \eta_{\mu\nu} + u_\mu u_\nu~~.
\end{equation}
where $x^\mu=(v,x_{i})$, velocities $\beta^i $ are constants, $P_{\mu \nu}$ is the projector onto spatial directions, and the indices in the boundary are raised and lowered with the Minkowsik metric $\eta_{\mu\nu}$. The metric~(\ref{rnboost}) describes the uniform boosted black brane moving at velocity $\beta^i $~\cite{Bhattacharyya:2008jc}.

\section{The derivative expansion to first order}
In the boosted black brane solution~(\ref{rnboost}), the parameters $b$ and $\beta^i$ are related with the temperature and velocity, respectively. If we let the above parameters $b$ and $\beta^i$ be slowly-varying functions of the boundary coordinates $x^\mu=(v,x_{i})$, the metric~(\ref{rnboost}) will be not a solution of the equations of motion~(\ref{eqs}) any more. However, we can add correction terms to it and then make the new metric as a solution. Before adding these correction terms, we first define the tensors
\begin{eqnarray}
\label{Tensors} W_{\mu \nu} = R_{\mu \nu } -\frac{1}{2}Rg_{\mu \nu}+\alpha H_{\mu \nu}-6 g_{\mu \nu}~.
\end{eqnarray}
where we have set $l=1$ and hereafter. Obviously, if we take the parameters $b$ and $\beta^i$ as functions of $x^\mu$ in~(\ref{rnboost}), the $W_{\mu \nu}$ will be nonzero and proportional to the derivatives of the parameter functions. Therefore, these terms can be considered as the source terms and denoted as $S_{\mu \nu }$. As mentioned above, we can add the correction terms to cancel these source terms.
According to~\cite{Bhattacharyya:2008jc,Hur:2008tq,Erdmenger:2008rm,Banerjee:2008th,Tan:2009yg}, due to the background metric~(\ref{rnboost}) preserving a spatial $SO(3)$ symmetry,
these correction terms can be made order by order in a derivative expansion. And after made some gauge fixed, the choice for the $n$-th correction terms can be
\begin{eqnarray}\label{correction}
{ds^{(n)}}^2 = \frac{ k^{(n)}(r)}{r^2}dv^2 + 2 h^{(n)} (r)dv dr + 2 \frac{j^{(n)}_i(r)}{r^2}dv dx^i + \frac{r^2}{\ell_c^2} \left(\alpha^{(n)}_{ij} -\frac{2}{3} h^{(n)}(r)\delta_{ij}\right)dx^i dx^j.
\end{eqnarray}
Note that, in order to implement the above procedure, we can first obtain the solution at the origin, and then obtain the globally solution by constructing a covariant metric. In our paper, we just consider the first order correction terms and first order solution. Thus the parameters expanded around $x^\mu=0$ are
\begin{equation}
\beta_i=\partial_{\mu} \beta_{i}|_{x^\mu=0} x^{\mu},~~~b=b(0)+\partial_{\mu} b|_{x^\mu=0} x^{\mu}. \label{Expand}
\end{equation}
where we have assumed $\beta^i(0)=0$. After inserting the metric (\ref{rnboost}) with~(\ref{Expand}) into $W_{\mu \nu }$, the first order source terms can be $S^{(1)}_{\mu \nu } = - W_{\mu \nu }$. As mentioned above, we can consider the correction terms in (\ref{correction}) for $n=1$ to cancel these source terms, thus $W_{\mu \nu } = (\text{effect from correction}) - S^{(1)}_{\mu \nu }$ vanish. The more details of the equations of these correction terms are seen in the appendix~\ref{A}. By solving them, we obtain the first order correction terms and the first derivative order solution.

Note that, there are some relationships between these equations
\begin{eqnarray}
 &&W_{vi} + r^2 f(r) W_{ri} =0 ~:~ S_{vi} + r^2 f(r) S_{ri} = 0, \notag \\
 &&W_{vv} + r^2 f(r)W_{vr} =0 ~:~ S_{vv} + r^2 f(r) S_{vr} = 0. \label{constraint}
 \end{eqnarray}
which can be considered as the constrain equations.
In our paper, after using the first order source terms in the Gauss-Bonnet gravity (also seen in the appendix~\ref{A}), we can further rewrite the constrain equations~(\ref{constraint}) as
\begin{eqnarray}
&&3 \partial _vb+b \partial _i\beta _i=0, \notag \\
&&\partial _ib+b \partial _v\beta _i=0. \label{constraint1}
\end{eqnarray}
In the later, we can see that these equations can be expressed as a covariant form. And they are nothing but the exact conservation equations of the zeroth order stress-energy tensor.

Although the equations of these first order correction terms in the appendix~\ref{A} are more complicated than those of Einstein gravity, we can solve them after some calculations. And the coefficients of the first order correction terms are
\begin{eqnarray}
&&h(r) = 0,~~j_i = r^3\partial_v\beta_i,~~ k(r) = \frac{2}{3}r^3 \partial_i\beta^i,\\\nn &&\alpha_{ij} =  \alpha(r)\left\{ (\partial_i \beta_j + \partial_j \beta_i )-\frac{2}{3} \delta_{ij}\partial_k \beta^k \right\},
\end{eqnarray}
where $\alpha(r)$ and its asymptotic expression are
\begin{eqnarray}
\alpha(r)= \int_{\infty }^{r}\frac{s^{3}-2\alpha s^{2}[s^{2}f(s)]^{^{\prime
}}-(1-8\alpha )b^{3}}{-s+2\alpha \lbrack s^{2}f(s)]^{^{\prime }}}\frac{1}{%
s^{4}f(s)}ds\approx \ell_c^2(\frac{1}{r}-\frac{b^3U}{4r^4})+O(\frac{1}{r})^5.
\end{eqnarray}
Therefore, after adding the correction terms, the first-order metric expanded in boundary derivatives about $x^{\mu}=0$ is given explicitly as
\begin{eqnarray}
ds^{2}&=&2dvdr-r^{2}f(b_{0},r)dv^{2}+r^{2}dx_{i}^{2}-2x^{\mu }\partial _{\mu }\beta
_{i}dx^{i}dr-2x^{\mu}r^{2}[\frac{1}{\ell_c^2}-f(b_{0},r)]\partial _{\mu }\beta _{i}dx^{i}dv \notag\\
& &-r^{2}x^{\mu }C(r)\partial _{\mu }bdv^{2}
+2r^2\alpha(r)\sigma_{ij}dx^idx^j+\frac{2}{3}r\partial_i\beta^idv^2+2r\partial_v\beta_idvdx^i,\label{GloballySolution1}
\end{eqnarray}
where
\begin{equation}
f(b_{0},r)=f(b(x^{\mu}),r)|_{x^{\mu}=0},~~C(r)=\frac{\partial f(b(x^{\mu}),r)}{\partial b}|_{x^{\mu}=0},~~\sigma_{ij}=\partial_{(i} \beta_{j)}-\frac{1}{3} \delta_{ij}\partial_k \beta^k.
\end{equation}
From which, the global first-order metric can be constructed in a covariant form
\begin{eqnarray}
ds^2&=&-r^2 f(b,r) u_{\mu } u_{\nu } dx^{\mu } dx^{\nu }-2 u_{\mu } dx^{\mu } dr+\frac{r^2}{\ell_c^2} P_{\mu  \nu } dx^{\mu } dx^{\nu }+\Bigg[\frac{2 r}{3} u_{\mu } u_{\nu } \partial _{\lambda }u^{\lambda }\notag\\& & - r u^{\lambda } \partial _{\lambda }\left(u_{\mu } u_{\nu }\right) +2 r^2 \alpha (r) \sigma _{\mu  \nu }\Bigg] dx^{\mu } dx^{\nu },\label{GloballySolution}
\end{eqnarray}
where $\sigma_{\mu \nu}$ is
\begin{eqnarray}
\sigma ^{\mu  \nu }\equiv \frac{1}{2} P^{\mu  \alpha } P^{\nu  \beta } \left(\partial _{\alpha }u_{\beta }+\partial _{\beta }u_{\alpha }\right)-\frac{1}{3} P^{\mu  \nu } \partial _{\alpha }u^{\alpha }.
\end{eqnarray}
and we have also taken it as a covariant expression.

On the other hand, we investigate the corresponding boundary stress-tensor in the Gauss-Bonnet gravity. For the $n=5$ dimensional spacetime in G-B gravity, the corresponding boundary stress-tensor can be obtained through the variation of the total action
with respect to the boundary metric $\gamma_{ab}$
\begin{equation}
T_{ab}=\frac{2}{\sqrt{-\gamma }}\frac{\delta }{\delta \gamma
^{ab}}\left( I+I_{\mathrm{sur}}+I_{\text{ct}}^0 \right),
\label{Tab}
\end{equation}
where
\begin{eqnarray}
\label{Lagrangianct} I_{\mathrm{ct}}^0 &=&\frac{1}{8\pi
G}\int_{\partial \mathcal{M}} d^{4}x\sqrt{-\gamma } \left[
  -\frac{2+U}{\ell_c }
 -\frac{\ell_c }{4}(2-U)\mathsf{R}\right].
\end{eqnarray}
is the corresponding boundary counterterm and  $\mathsf{R}$ is the
curvature scalar associated with the
induced metric on the boundary $\gamma_{ab}$~\cite{Brihaye:2008xu}. Note that, it is obviously seen that the above boundary counterterm can recover the known counterterm expression in the Einstein gravity when $\alpha \to 0$~\cite{Balasubramanian:1999re,Emparan:1999pm,Mann:1999pc}.

From~(\ref{Tab}), the boundary stress-energy tensor is
\begin{equation}
 T_{ab}=\frac{1}{8 \pi G}[K_{ab}-\gamma _{ab}K
 +2\alpha (Q_{ab}-\frac{1}{3}Q\gamma_{ab})
-\frac{2+U}{\ell_c}\gamma _{ab} +\frac{\ell_c}{2}(2-U)%
( \mathsf{R}_{ab}-\frac{1}{2}\gamma _{ab}\mathsf{R})], \label{TabCFT}
\end{equation}
where $Q_{ab}$ is
\begin{eqnarray}
&Q_{ab}= 2KK_{ac}K^c_b-2 K_{ac}K^{cd}K_{db}+K_{ab}(K_{cd}K^{cd}-K^2)
+2K \mathsf{R}_{ab}+\mathsf{R}K_{ab} -2K^{cd}\mathsf{ R}_{cadb}-4
\mathsf{R}_{ac}K^c_b.~{~~~~~~} \notag
\end{eqnarray}
Therefore, after inserting the explicit metric~(\ref{GloballySolution1}) into~(\ref{TabCFT}), the corresponding non-zero components of the boundary energy-momentum tensor are
\begin{equation} T_{vv}=\frac{3b^4}{16\pi G r^2\ell_{c}},~T_{ij}=-\frac{U^2 b^3}{8\pi G r^2 \ell_c} \sigma_{ij}~(i\neq j),
~T_{ij}=\frac{1}{16\pi G}[\frac{b^4}{r^2 \ell_c}-\frac{2U^2 b^3}{r^2 \ell_c} \sigma_{ij}]~(i=j). \label{BoundaryST}
\end{equation}

Now, we investigate the hydrodynamics of the dual conformal field on the boundary. The background metric upon which the dual field theory resides can be
$h_{ab}=\lim_{r \rightarrow \infty}
\frac{\ell^2_c}{r^2}\gamma_{ab}$, and the expectation value of the
stress tensor of the dual conformal theory $\tau _{ab}$ is computed using the relation~\cite{Myers:1999psa}
\begin{eqnarray}\label{relation}
\label{Tik-CFT} \sqrt{-h}h^{ab}<\tau _{bc}>=\lim_{r\rightarrow
\infty }\sqrt{-\gamma }\gamma ^{ab}T_{bc}.
\end{eqnarray}
where $T_{ab}$ is the boundary stress-energy tensor of the Gauss-Bonnet gravity.

Note that, according to~(\ref{Solution1}), the background metric $h_{ab}$ upon which the dual field theory resides is
\begin{equation}
ds^2=h_{ab}dx^adx^b=-dv^2+dx^2+dy^2+dz^2.
\end{equation}

In order to obtain a conformal Minkowsik metric, we can also make the coordinate transformation $v=u\ell_{c}, x^i=\ell_{c}x'^{i}$. Now, $h_{ab}$ becomes
\begin{equation}
ds^2=h_{ab}dx^adx^b=\ell_{c}^2(-du^2+dx'^{2}+dy'^{2}+dz'^{2}).\label{Newmetric2}
\end{equation}
and the corresponding boundary stress-energy tensor~(\ref{BoundaryST}) in these new coordinate system are
\begin{equation} T_{uu}=\frac{3b^4\ell_{c}}{16\pi Gr^2},~T_{i'j'}=-\frac{U^2\ell_{c}b^3}{8\pi Gr^2} \sigma_{ij}~(i'\neq j'),
~T_{i'j'}=\frac{b^4\ell_{c}}{16\pi Gr^2}-\frac{2U^2 b^3 \ell_{c}}{16\pi Gr^2} \sigma_{ij}~(i'=j'). \label{BoundaryST1}
\end{equation}

Therefore, according to~(\ref{relation}) and~(\ref{BoundaryST}), the expectation value of the
first order stress tensor of the dual theory $\tau _{\mu \nu}$ which is corresponding to the global metric~(\ref{GloballySolution}) can be
\begin{equation}
\tau_{\mu \nu}=\frac{1}{16\pi G}[\frac{b^4}{\ell_{c}^3}(\eta_{\mu\nu} + 4 u_\mu u_\nu)-\frac{2U^2b^3}{\ell_{c}^3}\sigma_{\mu\nu}]=P(\eta_{\mu\nu} + 4 u_\mu u_\nu  ) - 2 \eta \sigma_{\mu\nu}. \label{StressTensor}
\end{equation}
where we have rewritten it as a covariant form. It should be emphasized that we can also obtain~(\ref{StressTensor}) by directly inserting the global metric~(\ref{GloballySolution}) into~(\ref{TabCFT}). From~(\ref{StressTensor}), the pressure and viscosity are read off
\begin{equation}
P=\frac{b^4}{16 \pi G \ell_{c}^3},~~~\eta=\frac{U^2b^3}{16 \pi G \ell_{c}^3}=\frac{1}{16 \pi G \ell_{c}^3}(1-8\alpha)b^3. \label{ets}
\end{equation}
and the entropy density $s$ can be computed through
\begin{equation}
s=\frac{\partial P}{\partial T}=\frac{1}{4G \ell_{c}^3}\pi^3T^3=\frac{b^3}{4G \ell_{c}^3}. \label{entr1}
\end{equation}
Note that, because we have made a time coordinate transformation, thus the temperature in~(\ref{Newmetric2}) becomes $T=b\ell_{c}/(\pi)$. And
the temperature in~(\ref{entr1}) is $T=b/\pi$.
It is easily seen that (\ref{entr1}) is consistent with~(\ref{entr}).
And the shear viscosity is same as the result in reference~\cite{Brigante:2007nu}.
The ratio of (\ref{ets}) and (\ref{entr}) can also be found that
\begin{equation}
\label{ror}
 {\eta \over s} = {1 \over 4 \pi} (1-8\alpha).
\end{equation}
In addition, from $\tau _{\mu \nu}$, we can obtain the zeroth order energy-momentum tensor
\begin{equation}
\tau_{(0)}^{\mu\nu} =\frac{b^4}{16 \pi G \ell_{c}^3} (\eta^{\mu\nu} + 4 u^\mu u^\nu),
\end{equation}
and after a little work, the constrain equation~(\ref{constraint}) can be expressed covariantly as
\begin{equation}
\partial_\mu \tau_{(0)}^{\mu\nu} = 0 ~~.
\end{equation}
which is just the exact conservation equations of the zeroth order stress-energy tensor.


\section{Conclusion and discussion}
In this paper, we apply the AdS/CFT correspondence to investigate
the hydrodynamics of the dual conformal filed on the boundary in the
Gauss-Bonnet gravity. As an effective description of the conformal
field at long wavelengths, the stress-energy tensor of the
hydrodynamics is calculated to the first order. And from this first
order stress-energy tensor, we also calculate the shear viscosity
and energy density of the hydrodynamics. And these results are found
to be consistent with those of some previous works. Note that, in
principle, we can repeat the first order process to obtain the $n$th
order solution from the $(n-1)$ th order solution
and $n$th derivative order source terms, and then we can extract
more information of the strongly coupled conformal field by
calculating the corresponding stress-energy tensor. However, these
calculations must be more and more complicated with respect to the
order, and they would be taken into account in the future work. In addition, there are other methods to calculate the transport coefficients~\cite{Bigazzi:2010ku,Kanitscheider:2009as}, which can give us some new insights of the gravity/hydrodynamics. Therefore, it would be interesting to have further study on the underlying relevant relationships between these results.


\section{Acknowledgements}

Y.P Hu thanks Professor Rong-Gen Cai for his encouragement at the
first stage of this paper. And he also thanks Professor X.Jaen, Dr.
Hai-Qing Zhang, Dr. Li-Ming Cao for their useful discussions and
communication. Z.Y Nie thanks Dr. Hong-Bao Zhang and Xin Gao for
useful discussions. The calculation are mainly calculated by the TTC
package which is a tool of tensor calculus. And this work is
supported partially by grants from NSFC, China (No. 10773002, No.
10875018, No. 10873003, No. 10821504, No. 10975168, No. 11035008 and No. 11075098).

\appendix

\section{The tensor components of $W_{\mu \nu }$ and $S_{\mu \nu }$}
\label{A}
The tensor components of $W_{\mu \nu } = (\text{effect from correction}) - S_{\mu \nu }$ are
\begin{eqnarray}
W_{vv} &=&(-8r^{2}f^{2}h-\frac{6k}{r^{2}}+\frac{6fk}{r^{2}}
-7r^{3}fhf^{\prime}+\frac{3kf^{\prime }}{2r}-2r^{3}f^{2}h^{\prime}-r^{4}ff^{\prime}h^{\prime }+\frac{fk^{\prime}}{2r}
-r^{4}fhf^{\prime \prime}-\frac{fk^{\prime \prime }}{2}) \notag\\
&&+\alpha (432r^{2}f^{3}h-\frac{12f^{2}k}{r^{2}}+336rf^{2}hf^{\prime}-
\frac{12fkf^{\prime}}{r}+60r^{4}fhf^{\prime 2}+208rf^{3}h^{\prime}+90r^{4}f^{2}f^{\prime}h^{\prime}\notag\\
&&+9r^{5}ff^{\prime 2}h^{\prime}
+\frac{18f^{2}k^{\prime }}{r}+6ff^{\prime}k^{\prime}+24r^{4}f^{2}hf^{\prime \prime}+6r^{5}fhf^{\prime}f^{\prime
\prime}+20r^{4}f^{3}h^{\prime \prime }+6r^{5}f^{2}f^{\prime
}h^{\prime \prime }+12f^{2}k^{\prime \prime }\notag\\
&&+3rff^{\prime}k^{\prime \prime })
+(\frac{r^{2}f}{2}-12r^{2}\alpha f^{2}-3r^{3}\alpha ff^{\prime})R^{(1)}-S_{vv}, \notag\\
W_{vr} &=&\frac{1}{2r^{3}}(-12r^{3}h+28r^{3}fh+17r^{4}hf^{\prime}+4r^{4}fh^{\prime }+2r^{5}f^{\prime }h^{\prime
}+2r^{5}hf^{\prime \prime }-k^{\prime }+rk^{\prime \prime})+\alpha (-444f^{2}h \notag\\
&&+\frac{17}{2}rhf^{\prime }-342rfhf^{\prime }-60r^{2}hf^{\prime2}-208rf^{2}h^{\prime }-90r^{2}ff^{\prime }h^{\prime}-9r^{3}f^{\prime 2}h^{\prime }-24r^{2}fhf^{\prime \prime}-6r^{3}hf^{\prime}f^{\prime \prime } \notag\\
&&-20r^{2}f^{2}h^{\prime \prime }-6r^{3}ff^{\prime }h^{\prime \prime})+\frac{\alpha }{r^{3}}(6kf^{\prime }-18fk^{\prime }-6rf^{\prime
}k^{\prime}-12rfk^{\prime \prime }-3r^{2}f^{\prime }k^{\prime
\prime })+(-\frac{1}{2}+12\alpha f \notag\\
&&+3r\alpha f^{\prime })R^{(1)}-S_{vr}, \notag\\
W_{vi} &=&\frac{1}{2r}(3fj_{i}^{\prime }-rfj_{i}^{\prime \prime })-
\frac{\alpha f}{r}(8j_{i}f^{\prime }+6fj_{i}^{\prime }-2rf^{\prime
}j_{i}^{\prime }-2rfj_{i}^{\prime \prime })-S_{vi}, \notag\\
W_{rr} &=&\frac{1}{r^{5}}(1-4\alpha f)(r^{5}h^{\prime })^{\prime }-S_{rr}, \notag\\
W_{ri} &=&\frac{1}{2r^{3}}(-3j_{i}^{\prime }+rj_{i}^{\prime \prime})+\frac{\alpha }{r^{3}}(8j_{i}f^{\prime }+6fj_{i}^{\prime
}-2rf^{\prime }j_{i}^{\prime}-2rfj_{i}^{\prime \prime })-S_{ri}, \notag\\
W_{ij}\delta ^{ij} &=&\Big(\frac{3r^{2}}{2}(-1+24\alpha f+16r\alpha f^{\prime}+2r^{2}\alpha f^{\prime \prime })R^{(1)}-r^{2}h[-12+1272\alpha
f^{2}+452r^{2}\alpha f^{\prime 2}+r^{2}f^{\prime \prime }\notag\\
&&+6r^{4}\alpha f^{\prime \prime 2}+4f(-3+364r\alpha f^{\prime }+41r^{2}\alpha f^{\prime \prime})+2f^{\prime}(r+54r^{3}\alpha f^{\prime \prime})]-(-11r^{3}fh^{\prime }+644r^{3}\alpha f^{2}h^{\prime}\notag\\
&&-r^{4}f^{\prime }h^{\prime }
+530r^{4}\alpha ff^{\prime}h^{\prime }
+68r^{5}\alpha f^{\prime 2}h^{\prime }
-6\alpha kf^{\prime \prime }
+56r^{5}\alpha fh^{\prime }f^{\prime
\prime }+9r^{6}\alpha f^{\prime }h^{\prime }f^{\prime \prime}+6r\alpha k^{\prime}f^{\prime \prime }\notag\\
&&-r^{4}fh^{\prime \prime}
+64r^{4}\alpha f^{2}h^{\prime \prime }+44r^{5}\alpha ff^{\prime}h^{\prime \prime }+6r^{6}\alpha ff^{\prime \prime}h^{\prime
\prime }+30\alpha fk^{\prime \prime }+24r\alpha f^{\prime}k^{\prime \prime}+3r^{2}\alpha f^{\prime \prime }k^{\prime \prime})\notag\\
&&-\frac{-3k^{\prime}+72\alpha fk^{\prime }+36r\alpha f^{\prime }k^{\prime }}{r} \Big)
/\ell_{c}^2-S_{ij}\delta ^{ij}, \notag\\
W_{ij} &=&\left(\frac{1}{3}\delta _{ij}(\delta ^{kl}W_{kl})-\frac{1}{2r}
(r^{5}f(1-4\alpha f-2\alpha rf^{\prime })\alpha _{ij}^{\prime
})^{\prime} \right)/\ell_{c}^2-S_{ij}+\frac{1}{3}\delta _{ij}(\delta ^{kl}S_{kl}),
\end{eqnarray}
where
\begin{equation}
R^{(1)}=40 f h + 20 r h f^{\prime } + 20 r f h^{\prime} +3r^2f^{\prime} h^{\prime}+\frac{2k^{\prime}}{r^3}+2r^2hf^{\prime \prime}+2r^2fh^{\prime \prime}+\frac{k^{\prime \prime}}{r^2}.
\end{equation}
And the first order source terms are
\begin{eqnarray}
S_{vv} &=&(\frac{3rC}{2}-6\alpha rCf)\partial _{v}b-\frac{%
(2rf+r^{2}f^{\prime})\partial _{i}\beta_{i}}{2}+\alpha
(84rf^{2}+22r^{2}ff^{\prime})\partial _{i}\beta_{i} \notag\\
&&+\frac{8\partial _{i}\beta_{i}}{r}(\frac{r^{2}f}{2}-12r^{2}\alpha f^{2}-3r^{3}\alpha ff^{\prime}), \notag\\
S_{vr} &=&\frac{\partial _{i}\beta _{i}}{r}-\alpha (\frac{84f}{r}%
+20f^{\prime})\partial _{i}\beta _{i}+\frac{8\partial _{i}\beta _{i}}{r}(-\frac{1}{2}+12\alpha f+3r\alpha
f^{\prime }), \notag\\
S_{vi} &=&\frac{3rC+r^{2}C^{\prime}}{2}\partial _{i}b-\alpha
(6rCf+2r^{2}fC^{\prime }+2r^{2}Cf^{\prime })\partial _{i}b+\frac{
3rf+r^{2}f^{\prime}}{2}\partial _{v}\beta _{i}-\alpha
(6rf^{2}+4r^{2}ff^{\prime })\partial _{v}\beta _{i}, \notag\\
S_{rr} &=&0, \notag\\
S_{ri} &=&-\frac{3}{2r}\partial _{v}\beta _{i}+\alpha (\frac{6f}{r}%
+2f^{\prime })\partial _{v}\beta _{i}, \notag\\
S_{ij} &=&\left(\frac{3}{2}r(\partial _{i}\beta _{j}+\partial _{j}\beta
_{i})-\alpha (6rf+6r^{2}f^{\prime }+r^{3}f^{\prime \prime })(\partial
_{i}\beta _{j}+\partial _{j}\beta _{i})\text{ \ }(i\neq j)\right)/\ell_{c}^2, \notag\\
S_{xx} &=&\Big(3r\partial _{x}\beta _{x}-\alpha (12rf+12r^{2}f^{\prime}+2r^{3}f^{\prime \prime})\partial _{x}\beta _{x}+r\partial _{i}\beta
_{i}-\alpha (84rf+52r^{2}f^{\prime}+6r^{3}f^{\prime \prime })\partial
_{i}\beta _{i} \notag\\
&&+\frac{8\partial _{i}\beta _{i}}{r}(\frac{-r^2}{2}+12r^{2}\alpha f+8r^{3}\alpha f^{\prime}+r^{4}\alpha f^{\prime \prime }) \Big)/\ell_{c}^2,
\end{eqnarray}
where $S_{yy}$ (or $S_{zz}$) is just replaced the index $x$ in $S_{xx}$ into $y$ (or $z$), and
\begin{equation}
C(r)=\frac{\partial f(r)}{\partial b}.
\end{equation}
Note that, there are two identities related with $f(r)$
\begin{eqnarray}
4(1-f)+8\alpha f^{2}-rf^{\prime }+4r\alpha ff^{\prime } &=&0, \notag\\
-12(1-f)-24\alpha f^{2}+8rf^{\prime}-32r\alpha ff^{\prime }
&=&4r^{2}\alpha f^{\prime 2}-r^{2}f^{\prime \prime}+4r^{2}\alpha
ff^{\prime \prime}.
\end{eqnarray}
And the equations of the first order correction terms can be obtained from $W_{\mu \nu }=(\text{effect from correction})-S_{\mu \nu }=0$. Here, $S_{\mu \nu }$ is the first order source term. In addition, it can be easily seen that the above formalisms $W_{\mu \nu }$ and $S_{\mu \nu }$ can recover the corresponding formalisms in Einstein Gravity when $\alpha=0$.

\end{document}